\documentclass[twocolumn,aps,prl,showpacs,superscriptaddress,10pt] {revtex4-1}

	\usepackage{amsmath}%,amssymb} 
	\usepackage{makeidx}
	\usepackage{amsfonts}
    \usepackage{dsfont}
	\usepackage[ansinew]{inputenc}
	\usepackage[usenames,dvipsnames]{pstricks}
	\usepackage{subfigure}
	\usepackage{epsfig}
	\usepackage{pst-grad} % For gradients
	\usepackage{pst-plot} % For axes
	\usepackage[hyperindex]{hyperref}
	\usepackage{color}
\usepackage{wasysym}
	
\makeindex

\begin{document}

\title{Quantitative Measurement of Giant and Quantized Microwave Faraday Rotation}% Force line breaks with \\

\author{Vishnunarayanan Suresh}
\affiliation{D\'{e}partement de Physique et Institut Quantique, Universit\'{e} de Sherbrooke, Sherbrooke, Qu\'ebec, J1K 2R1, Canada}
 \author{Edouard Pinsolle} 
\affiliation{D\'{e}partement de Physique et Institut Quantique, Universit\'{e} de Sherbrooke, Sherbrooke, Qu\'ebec, J1K 2R1, Canada}
\author{Christian Lupien}
\affiliation{D\'{e}partement de Physique et Institut Quantique, Universit\'{e} de Sherbrooke, Sherbrooke, Qu\'ebec, J1K 2R1, Canada}
 \author{Talia J. Martz-Oberlander}
\affiliation{Department of Physics, McGill University, Montr\'eal, Qu\'ebec, H3A 2T8, Canada}
\author{Michael P. Lilly}
\affiliation{Center for Integrated Nanotechnologies, Sandia National Laboratories, Albuquerque, NM 87185, USA}
\author{John L. Reno}
\affiliation{Center for Integrated Nanotechnologies, Sandia National Laboratories, Albuquerque, NM 87185, USA}
\author{Guillaume Gervais}
\thanks{To whom correspondence should be addressed. E-mail: gervais@physics.mcgill.ca, thomas.szkopek@mcgill.ca, Bertrand.Reulet@USherbrooke.ca }%
\affiliation{Department of Physics, McGill University, Montr\'eal, Qu\'ebec, H3A 2T8, Canada}
\author{Thomas Szkopek}
\thanks{To whom correspondence should be addressed. E-mail: gervais@physics.mcgill.ca, thomas.szkopek@mcgill.ca, Bertrand.Reulet@USherbrooke.ca }%
\affiliation{Department of Electrical and Computer Engineering, McGill University, Montr\'eal, Qu\'ebec, H3A 0E9, Canada}
\author{Bertrand Reulet}
\thanks{To whom correspondence should be addressed. E-mail: gervais@physics.mcgill.ca, thomas.szkopek@mcgill.ca, Bertrand.Reulet@USherbrooke.ca }%
\affiliation{D\'{e}partement de Physique et Institut Quantique, Universit\'{e} de Sherbrooke, Sherbrooke, Qu\'ebec, J1K 2R1, Canada}

%\affiliation{$^{a}$D\'{e}partement de Physique et Institut Quantique, Universit\'{e} de Sherbrooke, Sherbrooke, Qu\'ebec, J1K 2R1, Canada}
%\affiliation{$^{b}$Department of Physics, McGill University, Montr\'eal, Qu\'ebec, H3A 2T8, Canada}
%\affiliation{$^{c}$Center for Integrated Nanotechnologies, Sandia National Laboratories, Albuquerque, NM 87185, USA}
%\affiliation{$^{d}$Department of Electrical and Computer Engineering, McGill University, Montr\'eal, Qu\'ebec, H3A 0E9, Canada}

\date{\today}% It is always \today, today,

\begin{abstract}

We report {\it quantitative} microwave Faraday rotation measurements conducted with a high-mobility two-dimensional electron gas (2DEG) in a GaAs/AlGaAs semiconductor heterostructure. In a magnetic field, the Hall effect and the Faraday effect arise from the action of Lorentz force on electrons in the 2DEG. As with the Hall effect, a classical Faraday effect is observed at low magnetic field as well as a quantized Faraday effect at high magnetic field. The high electron mobility of the 2DEG enables a giant single-pass Faraday rotation of $\theta_F^{max} \simeq 45^\circ$ $(\simeq0.8$~rad) to be achieved at a modest magnetic field of $B \simeq 100$~mT. In the quantum regime, we find that the Faraday rotation $\theta_F$ is quantized in units of $\alpha^*= 2.80(4)\alpha$, where $\alpha\simeq 1/137$ is the fine structure constant. The enhancement in rotation quantum $\alpha^* > \alpha$ is attributed to electromagnetic confinement within a waveguide structure.

\end{abstract}

\maketitle

{\bf Introduction.} Faraday rotation is the phenomenon whereby the polarization state of linearly polarized light is rotated by matter under the influence of a magnetic field applied along the direction of propagation \cite{Faraday46}. Faraday rotation manifests itself in a wide range of physical settings, from the passage of radio frequency waves through interstellar gas \cite{Smith68} to X-ray transmission through iron films \cite{Froman32}. Beyond electromagnetic waves alone, the acoustic analog of Faraday rotation has been used as a probe of the superfluid properties of $^3$He-B \cite{Halperin99}, wherein spin-orbit locking couples acoustic response with magnetic field.  In a semiconducting two-dimensional electron gas (2DEG), preliminary evidence of a quantized Faraday effect  in the microwave regime reminiscent of the quantum Hall effect was observed by Volkov and co-workers in 1986 \cite{Volkov85,Volkov86}. More recently, Faraday rotation has also been used in the THz domain as a probe of the topological properties of low-dimensional electron systems \cite{Shimano10, Crassee11, Shimano13, Pimenov16, Armitage16, Molenkamp17}. Here, we report on {\it quantitative} microwave measurements of Faraday rotation in a high-mobility 2DEG. A giant Faraday rotation of $\simeq0.8$~rad is observed, exceeding the previous record of giant Faraday rotation by eight-fold \cite{Crassee11}. In the quantum limit, the rotation angle is observed to be quantized at multiple filling factors of the integer quantum Hall effect in units of an effective fine structure constant $\alpha^*$ whose scale is set by the fine structure constant $\alpha\simeq 1/137$. \\

\begin{figure}
    \centering
    \includegraphics[width=.48\textwidth]{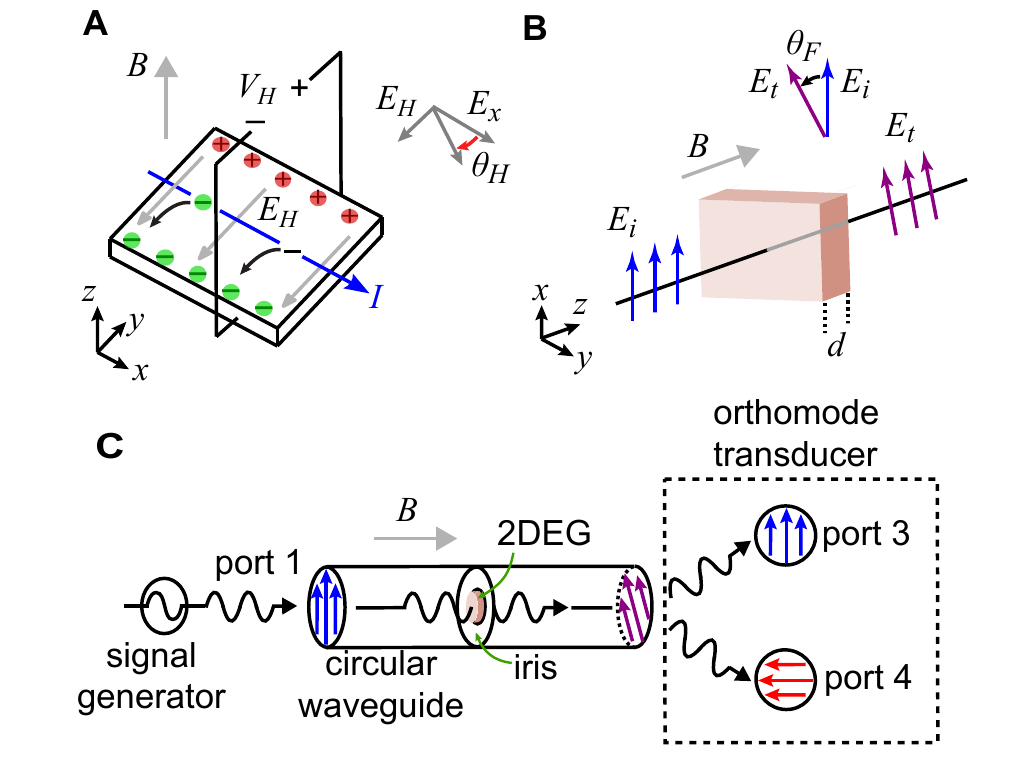}
    \caption{{\bf Classical Hall/Faraday effects and experimental setup}. A schematic representation of the classical Hall ({\bf A}) and Faraday effects ({\bf B}) is shown, along with the definition of the Hall angle $\theta_H$ and the Faraday rotation angle $\theta_F$. ({\bf C}) Experimental setup to measure microwave Faraday rotation. A linearly-polarized electromagnetic wave is injected into a circular hollow waveguide (port 1) that supports two orthogonally polarized TE$_{11}$ modes. The transmitted field is measured using an orthomode transducer in a direction parallel (port 3) and perpendicular (port 4) to the incoming electromagnetic wave.}
      \label{Fig1}
\end{figure}

 The Faraday and Hall effects in a 2DEG have a common origin with the cyclotron motion of charge carriers arising from the action of Lorentz force in the presence of an applied magnetic field $B$. As depicted in Fig.\ref{Fig1}{\bf (A)}, the Hall effect is the generation of an electric field $\vec{E}_{H}$ transverse to the direction of current flow $I$ and magnetic field $B$. The Hall effect is usually quantified by the transverse Hall resistivity $\rho_{xy}=V_{H}/I=B/ne$, where $n$ is the electron sheet density and $e$ the electric charge. In the classical regime, the Hall effect can be described with a Hall angle $\theta_H = \rho_{xy}/\rho_{xx}$, where $\rho_{xx}$ is the longitudinal resistivity of the 2DEG. Similarly, the Faraday effect depicted in Fig.\ref{Fig1}{\bf (B)} also arises from the action of Lorentz force upon charge, ultimately resulting in the rotation of polarization of a linearly polarized electromagnetic wave. The Faraday rotation $\theta_F$ is the angle of linear polarization rotation. In many materials, Faraday rotation is weak and well described by a linear relation $\theta_F=VdB$, where $V$ is the Verdet constant and $d$ the thickness of the medium. As we will show in this work, the high mobility 2DEG enables exceptionally large Faraday rotation. \\

Consider first a 2DEG in a strong magnetic field, which can give rise to the quantum Hall effect (QHE) wherein $\rho_{xy}$ is quantized in units of $h/e^2$, the resistance quantum \cite{Klitzing80}.  In the high magnetic field limit of the integer \cite{Klitzing80} (or fractional \cite{Tsui82}) quantum Hall regime, the longitudinal conductivity is $\sigma_{xx} = 0$ and the transverse conductivity is given by $\sigma_{yx} = i e^2/h$ where $i$ is the integer filling factor ($\nu$ in the fractional regime). The relation between 2DEG current density $\vec{J}(\omega)$ and electric field $\vec{E}(\omega)$ is thus determined by the conductivity tensor, 
\begin{equation}
\vec{J}(\omega)=\hat{\sigma}\vec{E}(\omega)=\left(\begin{array}{cc}0& -ie^2/h \\ +ie^2/h & 0 \end{array}\right)\vec{E}(\omega),
%\vec{J}(\omega)=\left(\begin{array}{cc}\sigma_{xx} & \sigma_{xy} \\\sigma_{yx} & \sigma_{yy}\end{array}\right)\vec{E}(\omega)=\left(\begin{array}{cc}0& e^2/h \\ -e^2/h & 0 \end{array}\right)\vec{E}(\omega), 
\end{equation}
where the frequency $\omega \ll \omega_c$ with $\omega_c = e B /m^*$ the cyclotron frequency.\\

Volkov and Mikhailov\cite{Volkov85} were first to consider the ideal scenario of a 2DEG in the QHE regime in vacuum, probed by a normally incident electromagnetic plane wave of frequency $\omega \ll \omega_c$. In this limit, the transmitted electromagnetic field $\vec{E}_{t}(\omega)$ has contributions from both the incident field $\vec{E}_{i}$ and the forward scattered field that is generated by the quantized transverse current density in the 2DEG. The Faraday rotation angle is predicted by simple Fresnel analysis to become quantized \cite{Volkov85},
\begin{equation}
\tan(\theta_F) =  i \frac{Z_0}{2}\frac{e^2}{h} = i \alpha.
\label{EqVolkov}
\end{equation}
where $Z_0$ is the impedance of free space, and the fine structure constant $\alpha = Z_0 e^2/2h$ here sets the natural scale for Faraday rotation\cite{Volkov85, MacDonald10}. The microwave frequency range (300~MHz$~\lesssim f \lesssim~$300~GHz) is particularly suitable for experiments attempting to realize this idealized scenario because the ``low-frequency'' limit $\omega \ll \omega_c$ can easily be achieved. Early experimental works consisted solely of measurements of cross-polarized transmitted microwave power in arbitrary units. Although they have shown inchoate quantization of transverse microwave transmission through 2DEGs \cite{Volkov86,Schlapp86}, to date there have been no quantitative measurements of microwave Faraday rotation in the QHE regime. \\

Interestingly, Faraday rotation is a \textit{2D bulk} probe of the quantum Hall state. In the QHE at integer filling factors $i$, charge transport experiments probe 1D edge currents, but it is important to recall that the \textit{2D bulk} transverse conductivity $\sigma_{xy}$ is quantized in the quantum Hall regime \cite{Thouless82}, and Faraday rotation explicitly probes the conductivity quantization of the bulk. As will be shown below, Faraday rotation of electromagnetic waves explicitly probes the quantization of bulk conductivity. Understanding the microwave Faraday rotation of the integer quantum Hall regime is an important step towards understanding Faraday rotation in the more complex fractional quantum Hall (FQH) regime \cite{Tsui82} hosted in ultra-high-mobility 2DEGs. The FQH states of a 2DEG are governed by incompressible Laughlin-like liquids, and perhaps host even more exotic quantum states such as the Moore-Read Pfaffian \cite{MR91}, for example.\\

%\textbf{Results and Discussion} \\

%{\it Semiconductor Heterostructure.} The AlGaAs/GaAs semiconductor sample is a modulation-doped quantum well with a well thickness of $d = 30$~nm grown at the Center of Integrated Nanotechnologies at Sandia National Laboratories (wafer VA0141). Two delta-doped layers with a density of $2\times10^{12}$cm$^{-2}$ are located symmetrically about the well at a setback distance of 55~ nm. The midpoint of the quantum well is located 100~nm underneath the surface of the $\ell = 0.55$~mm thick semiconductor. \\

{\bf Experimental setup.} The experimental apparatus is illustrated schematically in Fig.\ref{Fig1}{\bf (C)}, consisting of a circular hollow waveguide assembly designed for polarization sensitive microwave scattering measurements at cryogenic temperatures with a magnetic field oriented along the waveguide axis. The silver-plated hollow waveguide of diameter 23.825~mm supports two orthogonally-polarized TE$_{11}$ modes. A high-mobility 2DEG hosted in an AlGaAs/GaAs heterostructure grown by molecular beam epitaxy on a $\ell$ = 0.55~mm thick GaAs substrate with square dimensions 10~mm $\times$10~mm was inserted within the waveguide using a copper plate with a 9~mm diameter aperture functioning as a waveguide iris. The AlGaAs/GaAs semiconductor sample is a modulation-doped quantum well with a well thickness of $d = 30$~nm grown at the Center of Integrated Nanotechnologies at Sandia National Laboratories (wafer VA0141). Two delta-doped layers with a density of $2\times10^{12}$cm$^{-2}$ are located symmetrically about the well at a setback distance of 55~ nm. The midpoint of the quantum well is located 100~nm underneath the surface of the $\ell = 0.55$~mm thick semiconductor. \\

The mobility of the 2DEG was determined to be $\mu \simeq 1\times 10^{6}$~{cm}$^{2}$V$^{-1}$s$^{-1}$ by way of quasi-DC transport measurements at $T\simeq 20$~mK on a piece cut from the same wafer (during a separate cool down). The electronic density $n$ of the 2DEG was determined from the Landau level sequence observed in the Faraday rotation (see below), and found to be $2.08 (5) \times 10^{11}$ cm$^{-2}$. A coaxial-to-circular waveguide adapter (port 1) was used to excite the 2DEG with a linearly polarized TE$_{11}$ mode. The perpendicular (port 4) and parallel (port 3) polarized TE$_{11}$ mode fields were collected with an orthomode transducer, which consists of orthogonally polarized electric dipoles coupled to coaxial transmission lines. The entire assembly was thermally anchored to the cold plate of a dilution refrigerator with a base temperature of $\sim$7~mK. All temperatures quoted in this work correspond to the temperature of the mixing chamber of the dilution refrigerator. While the incident microwave illumination and/or imperfect thermalization will raise the temperature of the 2DEG electronic bath above that of the mixing chamber, our temperature dependence study of the Faraday rotation angle suggests the electrons are cooled down to at least $\sim$200~mK. Finally, a $\pm$6~T magnetic field was applied along the waveguide axis using a superconducting solenoid with the positive (+) direction aligned with the direction of propagation of the incident microwave. \\

\begin{figure}
    \centering
    \includegraphics[width=.48\textwidth]{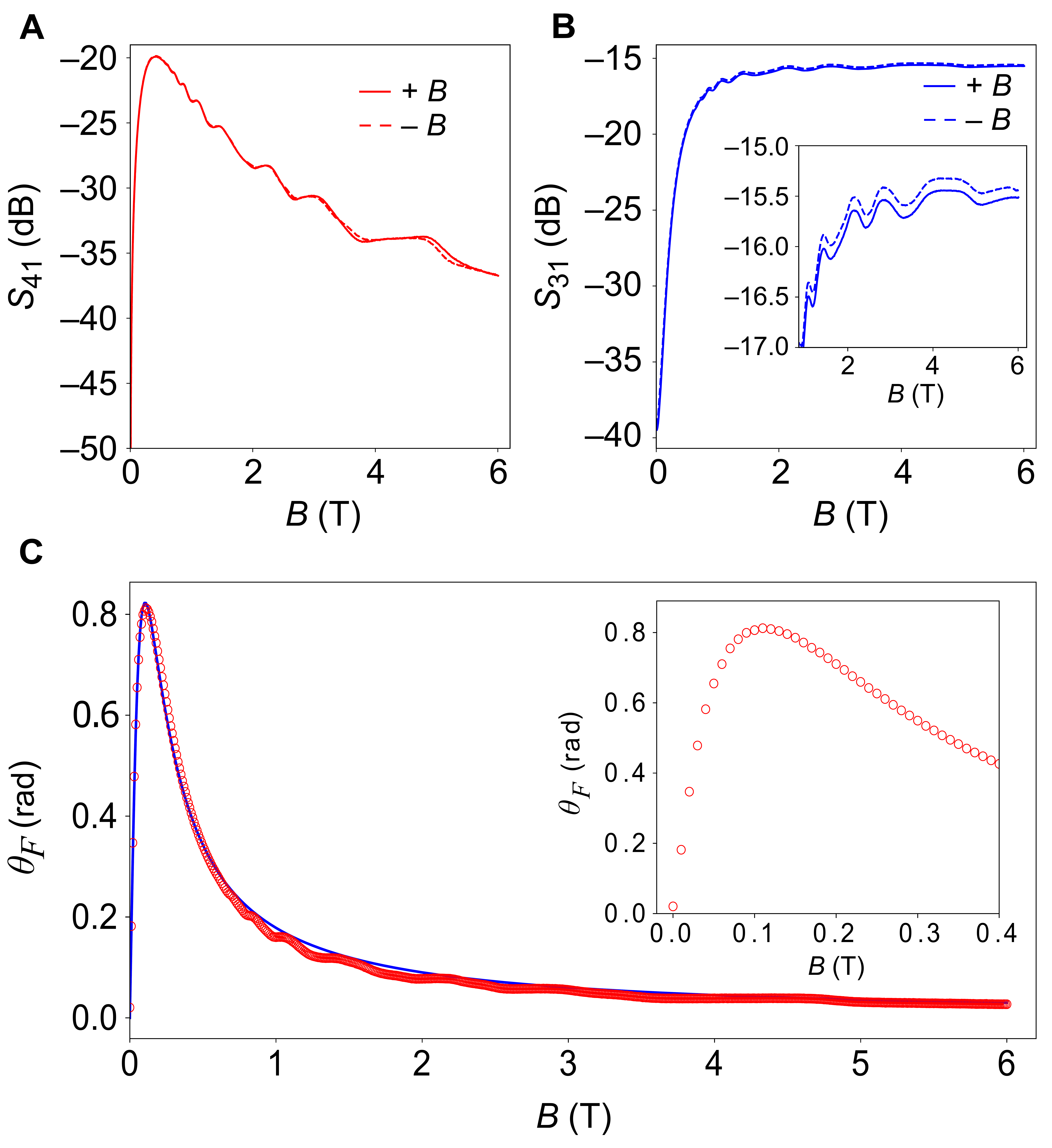}
    \caption{{\bf Scattering parameters and Faraday rotation measurements at 11.2GHz}. ({\bf A}) Perpendicular port scattering parameter $S_{41}$ and ({\bf B}) parallel port scattering parameter $S_{31}$ {\it versus} magnetic field $B$. The solid (dashed) line denotes the positive (negative) magnetic field  polarity. ({\bf C}) Magnetic field dependence of the Faraday angle $\theta_{F}$  (red circles) at the base temperature of the dilution refrigerator ($\sim$ 7 mK). The blue line is a fit of the Faraday rotation versus magnetic field with a classical Drude conductivity model (see text). The inset shows a zoom-in of the same data at low magnetic field. }
    \label{Fig2}
\end{figure}

The incident microwaves at 11.2 GHz were generated by a vector network analyzer (VNA) that was also used to measure the transmitted microwaves, thus enabling measurement of the scattering parameters (see Fig.\ref{Fig2}). High-frequency coaxial assemblies were used to couple the VNA to the hollow waveguide assembly in the dilution refrigerator. A low-temperature switch was used to transmit the microwaves from ports 3 and 4 of the hollow waveguide to the VNA using the same coaxial assembly, thereby limiting differences in transmission to the hollow waveguide apparatus. A cryogenic pre-amplifier was also used at the $\sim$3~K stage of the dilution refrigerator together with filters and attenuators to minimize microwave induced Joule heating of the 2DEG and suppress spurious reflections within the coaxial assembly.\\

{\bf Faraday Rotation Measurements.} The measured scattering parameter amplitudes $|S_{41}|$ and $|S_{31}|$ are shown in Fig. 2{\bf (A)} and {\bf (B)}  for perpendicular and parallel polarized transmission, respectively, versus applied magnetic field $B$. The  difference in the scattering parameter amplitudes of $\sim$0.1~dB  for positive and negative magnetic fields arises from a slight misalignment in excitation and detection ports. This corresponds to a systematic error of approximately $\sim$1\% in the field amplitude. The perpendicular polarization transmission amplitude $|S_{41}(B)|$ plotted versus $B$ in Fig. 2 {\bf (A)} reveals a staircase corresponding to quantization of perpendicularly polarized transmission related to Landau level formation in the 2DEG. \\

The magnetic field dependent Faraday rotation, $\theta_F (B)$, is determined from the scattering parameter amplitudes via $\tan( \theta_F(B) ) = |S_{41}(B)/S_{31}(B)|$. The Faraday rotation $\theta_F(B)$ is shown in Fig. 2 {\bf (C)} and a maximum Faraday rotation $\theta_{F}^{max} \simeq 45 ^\circ$ ($\simeq0.8$~rad) is observed at a modest applied magnetic field of $B\simeq100$~mT. This peak in $\theta_F$ demarcates the low magnetic field regime where $\theta_F$ increases with $B$ and the high-field regime where $\theta_F$ {\it decreases} with increasing $B$. \\

{\bf Electromagnetic confinement.} A quantitative model for the observed Faraday rotation can be arrived at by combining a simple theory for microwave transmission in a system with electromagnetic confinement, along with a Drude conductivity model for the 2DEG. It can be shown (see below) that Faraday rotation in a waveguide loaded with a 2DEG is generally given by
\begin{equation}
\tan(\theta_F) = \frac{\gamma Z \sigma_{yx}}{K+Z\sigma_{xx}}, 
\label{Rotation}
\end{equation}
where $Z$ is an effective wave impedance, $K$ an effective transmission coefficient, and $\gamma$ a mode coupling parameter. In the idealized free space scenario, $Z=Z_0$, $K=2$ and $\gamma =1$. A similar relation has been developed and applied to experiments for a simple hollow waveguide geometry without an iris\cite{Skulason15,Caloz12}. Notably, Eq.~\ref{Rotation} is general, applying even in the presence of an iris where the near-field distribution defies simple analytical solution \cite{Bethe,Novotny}. Electromagnetic confinement will generally cause $Z$, $K$ and $\gamma$ to deviate from their free space values.\\

We derive  Eq.~\ref{Rotation}  in the presence of electromagnetic confinement beginning with a linear response \textit{ansatz} for the transmitted (forward scattered) electric field $\vec{E}_{t}$, incident electric field $\vec{E}_{i}$, local electric field $\vec{E}_{loc}$ at the 2DEG and current density $\vec{J}$ in the 2DEG,
\begin{eqnarray}
\vec{E}_{loc} = \hat{K}_1 \vec{E}_{i} - \hat{Z}_1 \vec{J}, \\
\vec{E}_{t} = \hat{K}_2 \vec{E}_{i} - \hat{Z}_2 \vec{J},
\end{eqnarray}
where $\hat{K}_1$ and $\hat{Z}_1$ are linear operators giving the contributions to local electric field from the input field and current, respectively, and $\hat{K}_2$ and $\hat{Z}_2$ are linear operators giving the contributions to transmitted field from the incident field and current, respectively. The 2DEG current density $\vec{J} = \hat{\sigma} \vec{E}_{loc}$ where $\hat{\sigma}$ is the 2DEG conductivity tensor. The transmitted field can be expressed in two useful forms,
\begin{eqnarray}
\vec{E}_{t} &= \left[ \hat{K}_2 - \hat{Z}_2 \hat{Z}_1^{-1} \hat{K}_1 \right] \vec{E}_i + \hat{Z}_2 \hat{Z}_1^{-1} \vec{E}_{loc} \nonumber \\
%= \left[ \hat{K}_2 - \hat{Z}_2 \hat{Z}_1^{-1} \hat{K}_1 \right] \vec{E}_i - \hat{Z}_2 \hat{Z}_1^{-1} \left[ \hat{K}_1 \vec{E}_{i} - \hat{Z}_1 \vec{J} \right] \nonumber \\
&= \left[ \hat{K}_2 - \hat{Z}_2\hat{\sigma} \left( \mathds{1}+\hat{Z}_1\hat{\sigma} \right)^{-1} \hat{K}_1 \right] \vec{E}_{i}.
\end{eqnarray}
In the limit that the 2DEG is a perfect electric conductor with unbounded conductivity $|\hat{\sigma}| \rightarrow \infty$, the local electric field $\vec{E}_{loc} \rightarrow 0$ resulting in total reflection and null transmission $\vec{E}_t \rightarrow 0$. The operator identity follows,
\begin{equation}
0 = \hat{K}_2 - \hat{Z}_2 \hat{Z}_1^{-1} \hat{K}_1,
\end{equation}
and hence for arbitrary $\hat{\sigma}$ the incident and transmitted fields are related by,
\begin{equation}
\vec{E}_i = \hat{K}_1^{-1} \left( \mathds{1}+\hat{Z}_1\hat{\sigma} \right) \hat{Z}_1 \hat{Z}_2^{-1} \vec{E}_{t}.
\end{equation}
In a waveguide, the incident and transmitted far-fields are linear combinations of waveguide modes, and we restrict our attention to the scenario of two orthogonally polarized degenerate waveguide modes with all other modes cut-off (evanescent). Without loss of generality, the transmitted field is chosen to define the $x$-polarized mode,
\begin{equation}
\vec{E}_t = a_t \vec{\phi}_x(x,y),
\end{equation}
and the incident field is taken as a linear combination of the $x$-polarized and $y$-polarized modes,
\begin{equation}
\vec{E}_i = a_{ix} \vec{\phi}_x(x,y) + a_{iy} \vec{\phi}_y(x,y),
\end{equation}
with $a_t$, $a_{ix}$ and $a_{iy}$ the complex scalar amplitudes of transmitted and incident fields, and $\vec{\phi}_x(x,y)$, $\vec{\phi}_y(x,y)$ the $x$- and $y$- polarized mode field distributions in the $x,y$ plane transverse to the propagation axis $z$. Adopting a bra-ket notation for simplicity,
\begin{equation}
< u | \hat{A} | v > = \int \vec{\phi}^*_u(x,y) \cdot \hat{A} \vec{\phi}_v(x,y) dxdy,
\end{equation}
where $u,v \in \left\{x,y\right\}$. The Faraday rotation tangent defined in terms of mode amplitudes is,
\begin{eqnarray}
\tan(\theta_F) & = & \frac{a_{iy}}{a_{ix}} \nonumber \\
& = & \frac{ < y |  \hat{K}_1^{-1} \left( \mathds{1}+\hat{Z}_1\hat{\sigma} \right) \hat{Z}_1 \hat{Z}_2^{-1} | x > }{ < x |  \hat{K}_1^{-1} \left( \mathds{1}+\hat{Z}_1\hat{\sigma} \right) \hat{Z}_1 \hat{Z}_2^{-1} | x >, }
\end{eqnarray}
where $a_{ix}$ and $a_{iy}$ are determined by combining Eqs.~8-10 and taking inner products. In a system with axial symmetry about the $z$ axis, there is no cross-coupling between orthogonally polarized modes in the absence of a 2DEG, and it follows that:
\begin{eqnarray}
< y |  \hat{K}_1^{-1} \hat{Z}_1 \hat{Z}_2^{-1} | x > = 0.
\end{eqnarray}
The conductivity tensor $\hat{\sigma}$ of a 2DEG in a normally oriented static magnetic field has the structure,
\begin{equation}
\hat{\sigma} =
%\begin{pmatrix}
%  \sigma_{xx} & -\sigma_{yx} \\
%  \sigma_{yx} & \sigma_{xx}
% \end{pmatrix} \nonumber \\
\sigma_{xx} \left(\vec{x}\vec{x} + \vec{y}\vec{y}\right) + \sigma_{yx} \left(\vec{y}\vec{x} - \vec{x}\vec{y}\right),
\end{equation}
where dyadic vector notation is used. Assembling all of the above, the Faraday rotation is given by,
\begin{eqnarray}
\tan(\theta_F) & = & \frac{   < y |  \hat{K}_1^{-1} \hat{Z}_1\hat{\sigma} \hat{Z}_1 \hat{Z}_2^{-1} | x >  }{< x |  \hat{K}_2^{-1} + \hat{K}_1^{-1}\hat{Z}_1\hat{\sigma} \hat{Z}_1 \hat{Z}_2^{-1} | x >} \nonumber \\
& = & \frac{ \gamma Z \sigma_{yx} }{K + Z \sigma_{xx}},
\end{eqnarray}
where there are three scalar parameters that emerge,
\begin{eqnarray}
Z &=& < x | \hat{K}_1^{-1} \hat{Z}_1 \cdot \left(\vec{x}\vec{x} + \vec{y}\vec{y}\right) \cdot \hat{Z}_1 \hat{Z}_2^{-1} | x > \\
\gamma &=&  \frac{< y | \hat{K}_1^{-1} \hat{Z}_1 \cdot  \left(\vec{y}\vec{x} - \vec{x}\vec{y}\right) \cdot \hat{Z}_1 \hat{Z}_2^{-1} | x > }{ < x | \hat{K}_1^{-1} \hat{Z}_1 \cdot \left(\vec{x}\vec{x} + \vec{y}\vec{y}\right) \cdot \hat{Z}_1 \hat{Z}_2^{-1} | x > }  \\
K &=&  < x | \hat{K}_2^{-1} | x >,
\end{eqnarray}
whose values depend upon the detailed electric field distributions within the iris loaded waveguide.\\

{\bf Drude analysis.} We further approximate the 2DEG conductivity with a simple, classical Drude conductivity tensor,
\begin{equation}
\hat{\sigma}^{D}=\sigma_0\frac{1}{(1- i\omega\tau)^2 +(\omega_c\tau)^2}\left(\begin{array}{cc}1-i\omega\tau& -\omega_c\tau \\ \omega_c\tau & 1-i \omega\tau \end{array}\right),
%\vec{J}(\omega)=\left(\begin{array}{cc}\sigma_{xx} & \sigma_{xy} \\\sigma_{yx} & \sigma_{yy}\end{array}\right)\vec{E}(\omega)=\left(\begin{array}{cc}0& e^2/h \\ -e^2/h & 0 \end{array}\right)\vec{E}(\omega), 
\label{Drude}
\end{equation}
with $\sigma_0=ne^2\tau/m^{*}=ne\mu$ the Drude conductivity and $\omega_c$ the cyclotron frequency related to the charge carrier scattering time $\tau$ by $\omega_c \tau= \mu B$. The charge carrier scattering time deduced from mobility is $\tau = m^* \mu / e \simeq 38$~ps with $m^*=0.067m_{e}$ the effective mass in GaAs, and $\omega \tau \simeq 2.7$ for our experiment at $f$ = 11.2~GHz. The solid blue line of Fig. 2 {\bf (C)} shows a best fit of $\theta_F$ versus $B$ to the modulus of Eq.~\ref{Rotation} with the Drude conductivity model Eq.~\ref{Drude}. Two independent fit parameters associated solely with electromagnetic confinement were used, taking the values $\gamma =0.49$ and $Z/K=1192~\Omega$ for the optimized fit, with $Z/K$ assumed to be real for simplicity. \\

Notably, our simple model accurately captures the essential features of Faraday rotation $\theta_F$ versus $B$. In the low magnetic field regime, $\mu B \ll 1$, the rotation $\theta_F \approx \gamma \sigma_{yx}/\sigma_{xx} \propto B$, as observed in Fig. 2 {\bf (C)} for $B \ll 100$~mT. In the high magnetic field regime $\mu B \gg 1$, the rotation $\theta_F \approx \gamma (Z/K) \sigma_{yx} \propto 1/B$, as is coarsely observed in Fig. 2 {\bf (C)} for $B \gg 100$~mT. As shown below, analysis beyond a classical Drude model is required to describe Faraday rotation in the high-field regime. \\

{\bf Quantized Rotation.} The measured Faraday rotation angle tangent $\tan(\theta_F)$ is plotted versus $1/B$ (solid red line) in Fig. 3 {\bf (A)}. Six plateaus are clearly observed in $\tan(\theta_F)$ versus $1/B$ with the lowest three plateaus evenly spaced along both axes, and a further three evenly space plateaus are observed with twice the step-height. We confirm the origin of these Faraday rotation plateaus with the emergence of Landau levels by plotting a fan diagram of the assigned Landau level index $i$ for each plateau versus the reciprocal field $1/B$ of the mid-point of each plateau in Fig. 3 {\bf (B)}. The observed integer filling factor sequence $i$ = 2, 3, 4, 6, 8 follows the Landau level filling factor relation $i=nh/eB$ with an electron density $n=2.08 (5) \times 10^{11}$~cm$^{-2}$, consistent with quasi-DC transport studies performed on samples of the same semiconductor wafer hosting the 2DEG. Here, the expected spin degeneracy lifting of the Landau levels occurs in between integer filling $i$ = 4 and 6, at a magnetic field value $B\sim$1.8~T, again consistent with previous quasi-DC charge transport studies of 2DEGs hosted in similar heterostructures with comparable electron mobility and density. \\

The Faraday rotation was also measured during a separate cool down in a slightly different experimental configuration employing two coaxial assemblies. These measurements are shown in the inset of Fig. 3 {\bf (A)} with the temperature of the dilution refrigerator at $\sim$10~mK (red line) where quantization is visible, and at 3.2~K (blue line) where quantization is almost absent. In the quantum Hall regime, at temperatures $k_B T$ approaching the Landau level energy gap $\Delta$, thermal excitation of electrons across $\Delta$ gradually smears out conductivity quantization until it is ultimately absent. In our measurements, the plateaus of Faraday rotation $\theta_F$ cannot be resolved at 3.2~K, consistent with orbital quantization of the 2DEG by a strong magnetic field. \\

\begin{figure}[t]
    \centering
    \includegraphics[width=.48\textwidth]{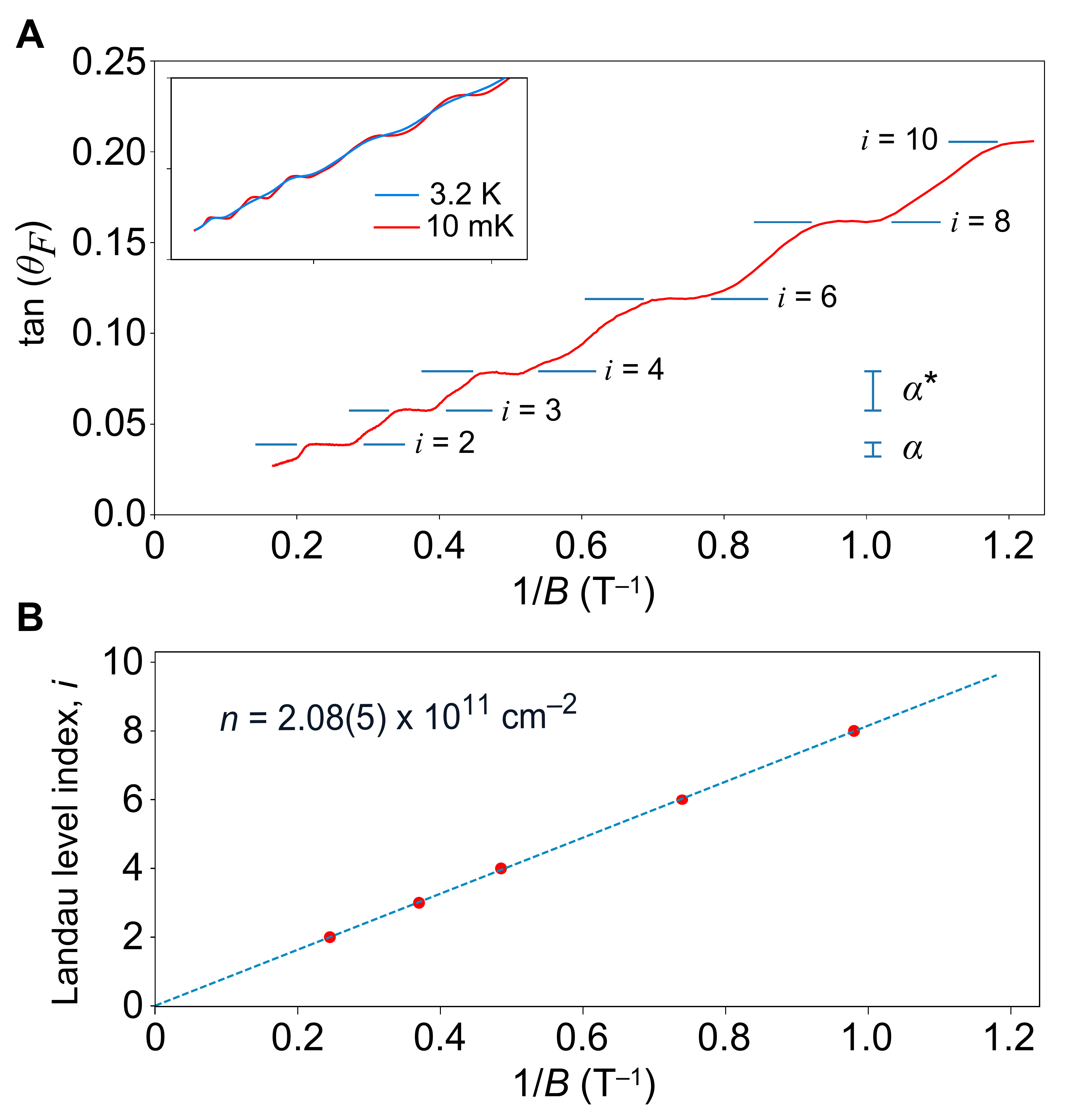}
    \caption{{\bf Quantized Faraday rotation}. ({\bf A}) Faraday angle plotted as $\tan(\theta_{F})$ {\it versus} $1/B$ (solid red line) at the base temperature of the dilution refrigerator. The expected position of each observed Faraday plateau is shown by horizontal markers with the quantization condition $\tan(\theta_{F})=i\alpha^*$.  The rotation quantum $\alpha$ for a 2DEG in vacuum is illustrated for reference. The inset shows a comparison of Faraday angle measurements at $\sim$10~mK (red line) and 3.2~K (blue line) temperature of the dilution refrigerator. ({\bf B}) Landau level index $i$ versus plateau mid-point $1/B$ (markers), with a linear fit (dashed line) from which the 2DEG electron sheet density $n$ is inferred. }
    \label{Fig3}
\end{figure}

Finally, we turn our attention to the observed value of quantized Faraday rotation. In the ideal free space scenario, the quantization condition is $\tan(\theta_F) = i \alpha$, with $\alpha$ the fine structure constant. The experimentally measured Faraday rotation of Fig. 3 {\bf (A)} exhibits a quantization $\tan(\theta_F)=i\alpha^*$. From a linear fit of the mid-points of each plateau in $\tan(\theta_F)$ versus $1/B$, the experimentally observed rotation quantum is $\alpha^* = 0.0204 (3) = 2.80(4)\alpha$. This is not surprising as the quantum of rotation in an ideal free-space scenario is $\alpha$, and electromagnetic confinement is expected to modify wave impedance and field distribution such that the rotation quantum in general differs from its free-space value, $\alpha^* \neq \alpha$. Applying our simple model, Eq.~\ref{Rotation}, for Faraday rotation to the QHE regime with $\sigma_{xx}=0$ and $\sigma_{yx}=ie^2/h$, rotation quantization takes a modified form,
\begin{equation}
\tan(\theta_F) =  i \frac{\gamma Z}{K} \frac{e^2}{h} = i \alpha^*,
\label{alphastar}
\end{equation}
where $\gamma$, $Z$ and $K$ are electromagnetic confinement parameters specific to the experimental geometry and frequency. The simple model estimate for the confinement enhanced rotation quantum using $\gamma=0.49$ and $Z/K=1192~\Omega$ as determined from the Drude model fit displayed in Fig. 2 {\bf (C)} is $\alpha^{*}=3.10\alpha$, agreeing within $10\%$ of the measured value $\alpha^{*}=2.80(4)\alpha$. \\

{\bf Conclusions.} We have measured the quantization of Faraday rotation in the quantum Hall regime in a high-mobility 2DEG. Microwave Faraday rotation plateaus are robust and well formed, allowing Landau level indexing and the observation of spin-splitting structure. Measurement of microwave Faraday rotation is thus a contactless method that may prove useful in probing low-dimensional electronic phenomena such as the quantum spin Hall effect \cite{Konig}, the quantum anomalous Hall effect \cite{Chang} and the fractional quantum Hall effect \cite{Tsui82}. Furthermore, as a consequence of the high mobilities achievable in the GaAs/AlGaAs 2DEG system, giant Faraday rotation reaching $\sim0.8$~rad can be obtained at modest applied magnetic fields of $\sim$100~mT. In the future, it is foreseeable that the Faraday effect arising from cyclotron motion of high mobility charge carriers in semiconductor materials and heterostructures could be used to {\it isolate} and {\it circulate} microwave signals, in lieu of conventional bulk ferrites that rely on off-resonant Larmor precession to impart Faraday rotation. \\

{\bf  Acknowledgements}. The authors thank R. C\^ot\'e and K. Bennaceur for helpful discussions, and S. Jezouin and G. Lalibert\'e for technical assistance. This work has been supported by Canada Excellence Research Chairs program, the Natural Sciences and Engineering Research Council of Canada, the Canadian Institute for Advanced Research, the Canadian Foundation for Innovation and the  Fonds de Recherche du Qu\'ebec Nature et Technologies. This work was performed, in part, at the Center for Integrated Nanotechnologies, an Office of Science User Facility operated for the U.S. Department of Energy (DOE) Office of Science. Sandia National Laboratories is a multimission laboratory managed and operated by National Technology and Engineering Solutions of Sandia, LLC, a wholly owned subsidiary of Honeywell International, Inc.,  for the U.S. DOE National Nuclear Security Administration under contract DE-NA-0003525. The views expressed in the article do not necessarily represent the views of the U.S. DOE or the United States Government.\\

%\textbf{Methods}

%\textbf{Semiconductor Heterostructure} The AlGaAs/GaAs semiconductor sample is a modulation-doped quantum well with a well thickness of $d = 30$~nm grown at the Center of Integrated Nanotechnologies at Sandia National Laboratories (wafer VA0141). Two delta-doped layers with a density of $2\times10^{12}$cm$^{-2}$ are located symmetrically about the well at a setback distance of 55~ nm. The midpoint of the quantum well is located 100~nm underneath the surface of the $\ell = 0.55$~mm thick semiconductor. \\

%\textbf{Theoretical Model} We derive a general formula for Faraday rotation in the presence of electromagnetic confinement beginning with a linear response \textit{ansatz} for the transmitted 

\bibliography{faraday}% Produces the bibliography via BibTeX.

\end{document}